\documentclass[preprintnumbers,superscriptaddress,amsmath,amssymb,prd,preprint,nofootinbib]{revtex4-2}
\usepackage{amssymb}
\usepackage{mathrsfs}
\usepackage{amsthm}
\usepackage{graphicx}
\usepackage{amsfonts}
\usepackage{ulem}
\usepackage[colorlinks,linkcolor=red,anchorcolor=blue,citecolor=green]{hyperref}
\usepackage{subfigure}
\usepackage{float}
\usepackage{enumerate}

\def\arctanh{\operatorname{arctanh}} 

\begin{document}	
\title{Revisiting critical orbits of test particles traveling in a black hole background\footnote{Submitted to Chinese Physics C}}
\author{Ping Li}
\email[]{Lip57120@huas.edu.cn}
\affiliation{College of Mathematics and Physics, Hunan University of Arts and Sciences, 3150 Dongting Dadao, Changde City, Hunan Province 415000, China}
\affiliation{Hunan Province Key Laboratory Integration and Optical Manufacturing Technology, 3150 Dongting Dadao, Changde City, Hunan Province 415000, China}
\author{Jun Cheng}
\email[]{chengjun@huas.edu.cn}
\affiliation{College of Mathematics and Physics, Hunan University of Arts and Sciences, 3150 Dongting Dadao, Changde City, Hunan Province 415000, China}
\affiliation{Hunan Province Key Laboratory Integration and Optical Manufacturing Technology, 3150 Dongting Dadao, Changde City, Hunan Province 415000, China}

\author{Jiang-he Yang}
\email[]{yjianghe@163.com}
\affiliation{College of Mathematics and Physics, Hunan University of Arts and Sciences, 3150 Dongting Dadao, Changde City, Hunan Province 415000, China}
\affiliation{Center for Astrophysics, Guangzhou University, 230 West Ring Road, Guangzhou, Guangdong Province 510006, China}

\begin{abstract}
This paper systematically revisits the critical orbits of test particles in various black hole backgrounds, including Schwarzschild, Reissner–Nordstr\"{o}m, Kerr, and Kerr–Newman spacetimes. We identify the critical orbits directly from the root structure of the radial equation, and we provide explicit expressions that relate the relevant parameters—energy, angular momentum, and charge‑to‑mass ratio—to the critical radius, as well as explicit formulas for the critical orbits in each case. Special attention is given to the relationships among the photon spheres, black hole shadows, and critical null geodesics. We also present extensive numerical results.
\end{abstract}
\maketitle

\section{Introduction}
Critical orbits are defined as those trajectories for which the radial kinetic energy function admits either a double or triple real root, corresponding to an unstable circular orbit or motion that neither falls into the black hole nor escapes to infinity. Critical orbits are powerful tools for analytically studying black hole shadows and accretion processes. However, as a specialized subclass of geodesics, they are rarely examined as stand-alone entities in the literature. In some literature, critical orbits may be overlooked in the broader discussion. The primary objective of this paper is to conduct a comprehensive review of the key properties of critical orbits, aiming to renew researchers' attention toward this significant topic in the field.

The study of geodesic motion in black hole spacetimes has a long and rich history, closely intertwined with the development of the solutions themselves. In spherically symmetric spacetimes, the earliest analytical solution for geodesic motion in Schwarzschild spacetime can be traced back to Droste \cite{Droste1917} in 1917, who expressed his solution in terms of the Weierstrass elliptic function. Thirteen years later, Hagihara conducted a comprehensive classification of all possible test particle motions in Schwarzschild spacetime, and his work \cite{Hagihara1931} has remained a classic reference. Simultaneously with the development of solutions based on Weierstrass functions, a significant body of research \cite{Forsyth1920,Greenhill1921,Darwin1959,Darwin1961,Rodriguez1987,Scharf2011} by Forsyth, Greenhill, Darwin, and Scharf \textit{et. al.} succeeded in expressing Schwarzschild geodesics using Jacobi elliptic functions and Legendre integrals, with publication years spanning several decades. In Reissner–Nordström (RN) spacetime, the motion of test particles exhibits richer behavior due to the presence of charge; for instance, a particle crossing the event horizon may not necessarily end at the singularity but can traverse the Cauchy horizon and emerge into another universe \cite{Grunau2011}. Grunau and Kagramanova \cite{Grunau2011} provided an analytical solution for the motion of electrically and magnetically charged test particles in this background.

In rotating, axisymmetric spacetimes, the theory of Kerr geodesics developed rapidly after Carter's fundamental discovery in 1968 that the Hamilton–Jacobi equation is completely separable \cite{Carter1968,Carter1968-1}. In this work, he identified a conserved quantity now universally known as the Carter constant, the geometrical nature of which was later understood by Walker and Penrose in terms of Killing tensors \cite{Walker:1970un}. Early analyses focused on equatorial orbits \cite{Felice1968,Bardeen1970-1,Bardeen1972-1}, while non-equatorial geodesics proved to be much more involved. Wilkins \cite{Wilkins1972} provided an early account of bound orbits, and vortical orbits were first discussed by De Felice and Calvani \cite{Felice1972,Felice1978}. A major technical breakthrough came in 2003, when Mino \cite{Mino2003} introduced the so-called Mino time, a parameter that completely decouples the radial and latitudinal equations of motion, enabling a straightforward application of elliptic functions. This facilitated the derivation of analytic solutions for bound timelike orbits by Fujita and Hikida \cite{Fujita2009}. The phase-space structure of Kerr geodesics, including homoclinic orbits and the separatrix, has been investigated by Levin and collaborators \cite{Levin2009,Perez2009}. More recently, null geodesics in the Kerr exterior have been revisited by Gralla and Lupsasca \cite{Gralla2020,Gralla2020-1}, providing convenient classifications and analytic solutions relevant to observational problems like black hole lensing and photon rings. And  Cie\'{s}lik \textit{et. al.} \cite{Cieslik2023} derive analytical solutions describing timelike and null geodesics by using Weierstrass elliptic functions in the Kerr spacetime. For the Kerr–Newman spacetime, which incorporates charge, the geodesic motion of both neutral and charged particles has been extensively studied. Early work by Johnston and Ruffini \cite{Johnston1974} examined timelike equatorial and spherical orbits of uncharged particles, while Young \cite{Young1976} analyzed the last stable orbit for charged particles. Bičák et al. \cite{Bicak1989-1,Bicak1989-2} provided a systematic study of charged particle motion, including radial motion and motion along the symmetry axis. Later, Kovář et al. \cite{Kovar2008} discovered unstable off-equatorial circular orbits for charged particles, and Pugliese et al. \cite{Pugliese2013} used equatorial circular orbits to distinguish between black holes and naked singularities. Hackmann and  Xu \cite{Hackmann2013} completely classifies the colatitudinal and radial motion of charged test particles in the Kerr-Newman spacetime and presents analytical solutions in terms of elliptic functions that are valid for all types of orbits. A comprehensive analysis of photon orbits in Kerr–Newman spacetime was presented by Calvani and Turolla \cite{Calvani1981}. For more recent literature, see \cite{Wang2017,Wang2022,Galtsov2019,Geoffrey2022,Chen2025}.

In the case of spacetimes with a cosmological constant, Hackmann and L\"{a}mmerzahl achieved a significant breakthrough by performing analytical integrations for Schwarzschild–de Sitter \cite{Hackmann2008-1,Hackmann2008-2} and Reissner–Nordstr\"{o}m–de Sitter geometries \cite{Hackmann2008-3} using hyperelliptic $\theta$ and $\sigma$ functions, based on the Jacobi inversion problem restricted to the $\theta$-divisor. This method was subsequently extended to higher-dimensional Schwarzschild and Reissner–Nordstr\"{o}m spacetimes with a cosmological constant \cite{Hackmann2008-3}. Similar hyperelliptic function methods have also been successfully applied to the axially symmetric Taub-NUT \cite{Kagramanova2010} and Kerr–de Sitter \cite{Hackmann2010} spacetimes, where the types of orbits are classified and extensively studied. In addition to the literature mentioned above, a number of books also discuss geodesic motion in black hole spacetimes in considerable detail, e.g. \cite{Bardeen1973,Chandrasekhar1983,Neil1995,Lammerzahl2016-1}.

Another important goal of this paper is to lay the groundwork for a 3+1 formalism describing the accretion of a Vlasov gas onto a Kerr–Newman black hole. In 2017, Rioseco and Sarbach \cite{Rioseco2017-1} developed a modern analytical theory for relativistic collisionless Vlasov gas accretion onto a Schwarzschild black hole. In their model, the tangential pressure exceeds the radial pressure at the horizon \cite{Rioseco2017-2}, offering a partial explanation for the issue of low accretion rates, which has attracted considerable attention. Subsequently, this model was quickly extended to describe accretion onto a moving Schwarzschild \cite{Mach2021-1,Mach2021-2}, Reissner–Nordstr\"{o}m \cite{Cieslik2020} and Kerr \cite{Li2023,Mach2025-1,Mach2025-2} black holes. In a related development, Mach et al. constructed a numerical framework using Monte Carlo methods to simulate the motion of collisionless gas in a Schwarzschild background \cite{Cieslik2023-11,Cieslik2024}. Since our focus is on accretion models, we are particularly interested in the scenario where test particles fall toward the black hole from infinity. Similar studies of unbound orbits in Schwarzschild spacetime have already been discussed, see \cite{Cieslik2022-1,Cieslik2023-1}. It should be noted that the accretion of neutral particles onto a black hole does not induce charge evolution. To investigate the effect of accretion on black hole charge, we have recently extended this model to describe the accretion of a collisionless Fermi gas onto a Reissner–Nordstr\"{o}m black hole \cite{Li2025}.

In the accretion model proposed by Rioseco and Sarbach \cite{Rioseco2017-1}, test particles are assumed to be in thermal equilibrium at infinity. Under gravitational influence, these particles follow geodesics into the finite region. Consequently, they naturally separate into three categories: those that fall into the black hole and are absorbed, those that are scattered by the black hole and return to infinity, and—lying between these two scenarios—a critical case in which particles neither enter the black hole nor escape to infinity, but instead asymptotically approach an unstable bound orbit. Therefore, examining this critical behavior is essential in accretion theory, as it defines the boundary in parameter space between absorbed and scattered particles. A similar line of reasoning applies to the analysis of the shadow boundary of a black hole. Therefore, this paper provides a systematic review of methods for handling critical geodesics in both spherically symmetric and axisymmetric rotating spacetimes. It aims to enable efficient identification of the relevant parameters associated with critical geodesics and to facilitate their analytical or numerical analysis.

In this paper, we systematically review the classical results related to critical geodesics for massless and neutral massive particles in the double root case, and present the relevant results for the triple root case and for charged particles in the Reissner–Nordstr\"{o}m and Kerr–Newman backgrounds. And the paper is organized as follows. In section II, we review the derivation of the equations of motion for charged test particles in Kerr–Newman spacetime using the separability of the Hamilton–Jacobi equation. Sections III through VI present a detailed analytical treatment of critical orbits — both null and timelike — in Schwarzschild, Reissner–Nordstr\"{o}m, Kerr, and Kerr–Newman black hole spacetimes, respectively. The final section provides conclusions and discussion. 

\section{The general equations of motion}
In this section, we review the equations of motion for charged particles moving in the Kerr-Newman spacetime. We work in the Boyer-Lindquist coordinates $(t,r,\theta,\varphi)$. The Kerr-Newman metric is well known and can be written as 
\begin{align}
    ds^2&=-\frac{\Delta-a^2\sin^2\theta }{\rho^2}dt^2-2\frac{a\sin^2\theta(2M r-Q^2) }{\rho^2}dtd\varphi +\frac{\rho^2}{\Delta} dr^2+\rho^2d\theta^2\nonumber\\
    &+\frac{(r^2+a^2)^2-\Delta a^2\sin^2\theta}{\rho^2}\sin^2\theta d\varphi^2, \label{KNm}
\end{align}
where
\begin{align}
    \Delta&=r^2-2Mr+a^2+Q^2,\\
    \rho^2&=r^2+a^2\cos^2\theta,
\end{align}
and $M$ is the  mass, $Q$ is the charge, and $a=J/M$ is the angular momentum per unit mass of the black hole. The outer and inner horizons are the roots of $\Delta(r)=0$ and are given by
\begin{equation}
    r_{\pm}=M\pm \sqrt{M^2-(a^2+Q^2)}.
\end{equation}
In general, the condition for a black hole to possess an event horizon is given by $0\leq a^2+Q^2\leq M^2$. The electromagnetic four-pontential in this background is $A_{\mu}= (-\frac{Q r}{\rho^2},\frac{Q r}{\rho^2}a\sin^2\theta,0,0)$. The inverse metric can be expressed as 
\begin{equation}
    g^{\mu\nu}=\left(\begin{array}{cccc}
        -\frac{ (r^2+a^2)^2-\Delta a^2\sin^2\theta}{\rho^2\Delta} &0 &0 &-\frac{a(2Mr-Q^2)}{\rho^2\Delta}\\
        0& \frac{\Delta}{\rho^2} &0&0\\
        0&0&\frac{1}{\rho^2}&0\\
        -\frac{a(2Mr-Q^2)}{\rho^2\Delta}&0&0 &\frac{\Delta-a^2\sin^2\theta}{\rho^2\Delta \sin^2\theta} 
     \end{array}\right),
\end{equation}
which satisfies $g^{\mu \alpha}g_{\alpha \nu}=\delta^{\mu}_{\nu}$.

Consider a test particle with mass $m$ and charge $q$ moving in the background of a rotating, charged black hole, the Lagrangian for its motion can be written as
\begin{align}
    \mathcal{L}& =\frac{1}{2}g_{\mu\nu}\frac{dx^{\mu}}{d\lambda}\frac{dx^{\nu}}{d\lambda}+e A_{\mu}\frac{dx^{\mu}}{d\lambda}\nonumber\\
    &=-\frac{1}{2}\frac{\Delta-a^2\sin^2\theta }{\rho^2}\Dot{t}^2-\frac{a\sin^2\theta(2M r-Q^2) }{\rho^2}\Dot{t}\Dot{\varphi} +\frac{1}{2}\frac{\rho^2}{\Delta} \Dot{r}^2+\frac{1}{2}\rho^2\Dot{\theta}^2\nonumber\\
    &+\frac{1}{2}\frac{(r^2+a^2)^2-\Delta a^2\sin^2\theta}{\rho^2}\sin^2\theta \Dot{\varphi}^2-\frac{eQr}{\rho^2}\Dot{t}+\frac{e Q r a\sin^2\theta}{\rho^2}\Dot{\varphi} ,
\end{align}
where $e=\frac{q}{m}$ is the charge-to-mass ratio, $\lambda$ is the affine parameter, and the dot denotes differentiation with
respect to $\lambda$. The canonical momenta $p_{\mu}$ are defined by
\begin{align}
    p_{t}&=-\frac{\partial \mathcal{L}}{\partial \Dot{t}}= \frac{\Delta-a^2\sin^2\theta }{\rho^2}\Dot{t}+\frac{2M r-Q^2 }{\rho^2}a\sin^2\theta\Dot{\varphi}+\frac{eQr}{\rho^2},\\
    p_{r}&=\frac{\partial \mathcal{L}}{\partial \Dot{r}}= \frac{\rho^2}{\Delta}\Dot{r},\\
    p_{\theta}&=\frac{\partial \mathcal{L}}{\partial \Dot{\theta}}=\rho^2 \Dot{\theta},\\
    p_{\varphi}&=\frac{\partial \mathcal{L}}{\partial \Dot{\varphi}}=
    \frac{(r^2+a^2)^2-\Delta a^2\sin^2\theta}{\rho^2}\sin^2\theta \Dot{\varphi}-\frac{2M r-Q^2 }{\rho^2}a\sin^2\theta\Dot{t}+\frac{e Q r a\sin^2\theta}{\rho^2}.
\end{align}
The Hamiltonian is obtained by performing a Legendre transformation 
\begin{equation}
    \mathcal{H}=-p_t\Dot{t}+p_r\Dot{r}+p_{\theta}\Dot{\theta}+p_{\varphi}\Dot{\varphi}-\mathcal{L}=\frac{1}{2}g_{\mu\nu}\frac{d x^{\mu}}{d\lambda}\frac{d x^{\nu}}{d\lambda}.
\end{equation}
There are four constants of motion: the rest mass $m$, the energy $E$, the Carter constant $D$, and the angular momentum $L_z$ in the $z$-direction, which are defined by
\begin{align}
    m^2&=-2\mathcal{H} =-g_{\mu\nu}\frac{d x^{\mu}}{d\lambda}\frac{d x^{\nu}}{d\lambda},\\
    E&=p_t=\frac{\Delta-a^2\sin^2\theta }{\rho^2}\Dot{t}+\frac{2M r-Q^2 }{\rho^2}a\sin^2\theta\Dot{\varphi}+\frac{eQr}{\rho^2},\\
    L_z&=p_{\varphi}=\frac{(r^2+a^2)^2-\Delta a^2\sin^2\theta}{\rho^2}\sin^2\theta \Dot{\varphi}-\frac{2M r-Q^2 }{\rho^2}a\sin^2\theta\Dot{t}+\frac{e Q r a\sin^2\theta}{\rho^2},\\
    D&=\rho^4\Dot{\theta}^2-\cos^2\theta \big(a^2(E^2-m^2)-\frac{L_z^2}{\sin^2\theta} \big).
\end{align}
With these constants, the equations of motion $\Dot{x}^{\mu}=\frac{\partial \mathcal{H}}{\partial p_{\mu}},\Dot{p}_{\mu}=-\frac{\partial \mathcal{H}}{\partial p_{\mu}}$ can be rewritten as a set of decoupled, first-order equations
\begin{align}
    \rho^2\Dot{t}&=\frac{1}{\Delta}\left[E((r^2+a^2)^2-\Delta a^2\sin^2\theta)+aL_z(Q^2-2Mr)-eQr(r^2+a^2) \right],\label{eom1}\\
    \rho^2\Dot{\varphi}&=\frac{1}{\Delta}\left[a E(2Mr-Q^2)+ \frac{L_z}{\sin^2\theta}(\Delta-a^2\sin^2\theta)-aQer\right],\\
    \rho^4\Dot{\theta}^2&\equiv\Theta  =D+\cos^2\theta \big(a^2(E^2-m^2)-\frac{L_z^2}{\sin^2\theta} \big),\label{THETA} \\
     \rho^4\Dot{r}^2&\equiv R=(E^2-m^2)r^4+2(m^2M-eQE)r^3+[a^2(E^2-m^2) +(e^2-m^2)Q^2-L_z^2-D]r^2\nonumber\\
     &+2[M((L_z-aE)^2+D)+aQe(L_z-aE)]r-Q^2[(L_z-aE)^2+D]-a^2D, \label{R} 
\end{align}
where we have introduced the variables $R=\rho^4\Dot{r}^2$ and $\Theta=\rho^4\Dot{\theta}^2$ for brevity.

We now introduce Hamilton's principal function $S$ via
\begin{equation}
    g_{\mu\nu}\Dot{x}^{\nu}+qA_{\mu}=\frac{\partial S}{\partial x^{\mu}}\equiv p_{\mu},
\end{equation}
and the Hamilton–Jacobi equation for $S$
 then takes the form
\begin{equation}\label{HmJac}
    -2\frac{\partial S}{\partial \lambda}=g^{\mu\nu}\left(\frac{\partial S}{\partial x^{\mu}}-qA_{\mu}\right)\left(\frac{\partial S}{\partial x^{\nu}}-qA_{\nu}\right).
\end{equation}
By demonstrating that the Hamilton–Jacobi equation (\ref{HmJac}) for test particles is separable, Carter \cite{Carter1968} first established the existence of a fourth constant of geodesic motion in Kerr spacetime -- now known as the Carter constant $D$. As a consequence of this separability, the abbreviated action $S$ is constrained to take a particular form
\begin{equation}
    S=\frac{1}{2}m^2\lambda-Et+L_z\varphi+\int^{r}\frac{\sqrt{R}}{\Delta}dr+\int^{\theta}\sqrt{\Theta}d\theta.\label{HmiJac}
\end{equation} 
Beacause $D$ is constant, generic geodesic motion is not planar; thus, the orbits lie in the three-dimensional coordinate space $(r,\theta,\varphi)$. The equations of motion are obtained by requiring that the partial derivatives of the abbreviated action $S$ with respect to the four constants $D,E,m,L_z$ vanish. Setting $\frac{\partial S}{\partial D}=0$, one yields the constraint satisfied by the $(r,\theta)$ coordinates
\begin{equation}\label{constraint}
    \int^r\frac{dr}{\sqrt{R}}=\int^{\theta}\frac{d\theta}{\sqrt{\Theta}}.
\end{equation}
Similarly, by setting$\frac{\partial S}{\partial m}=0$, $\frac{\partial S}{\partial E}=0$, and $\frac{\partial S}{\partial L_z}=0$, we find
\begin{align}
 \lambda&=\int^r\frac{r^2}{\sqrt{R}}dr+a^2\int^{\theta}\frac{\cos^2\theta } {\sqrt{\Theta}} d\theta,\\
    t&=E\lambda+\int^r\bigg((2ME-eQ)r^3-EQ^2r^2-a[2M(L_z-aE)+aeQ]r\nonumber\\
    &+aQ^2(L_z-aE)\bigg)\frac{dr}{\Delta\sqrt{R}},\\
    \varphi&=a\int^r[(2ME-eQ)r-L_za-EQ^2]\frac{dr}{\Delta\sqrt{R}}+L_z\int^{\theta} \csc^2\theta\frac{d\theta}{\sqrt{\Theta}}.\label{solphi}
\end{align}
The above expressions are obtained by simplification using Eq.(\ref{constraint}).

Consider test particles incident from infinity. Some of these particles will fall into the black hole, whereas others will be scattered back to infinity. Between these two cases lies a critical regime in which particles are neither captured by the black hole nor do they escape to infinity; instead, they asymptotically approach a critical radius at a finite distance from the black hole. We next provide a detailed analysis of the critical orbital solutions for test-particle motion, including both null and timelike orbits, in Schwarzschild, Reissner-Nordstr\"{o}m, Kerr, and Kerr-Newman spacetimes.
\section{Neutral particles traveling in a Schwarzschild space-time}
In this section, we consider the motion of neutral particles around a Schwarzschild black hole. In this case, the parameters satisfy $a=Q=e=0$. The equations of motion (\ref{eom1}) - (\ref{R}) reduce to 
\begin{align}
  \Dot{t}&=\frac{r^2}{\Delta}E,\\
  \Dot{\varphi}&=\frac{L_z}{r^2\sin^2\theta},\label{Seom1}\\
  \Dot{\theta}^2&=\frac{1}{r^4}\left(L^2-\frac{L_z^2}{\sin^2\theta} \right),\\
  r^4\Dot{r}^2&=(E^2-m^2)r^4+2m^2M r^3-L^2\Delta,\label{SR}
\end{align}
where $L^2=D+L_z^2$ denotes the squared angular momentum. In spherically symmetric spacetimes, the Carter constant $D$ is related to the conserved angular momentum $L^2$, which can be further expressed as
\begin{equation}
    L^2=r^4(\Dot{\theta}^2+\sin^2\theta \Dot{\varphi}^2 )\equiv r^4\frac{d\Omega^2}{d\lambda^2},
\end{equation}
where $d\Omega^2=d\theta^2+\sin^2\theta d\varphi^2$. When the square of the angular momentum $L^2$ is constant, it follows that the geodesics in this spacetime are confined to a single plane. This conclusion also holds for any spherically symmetric spacetime. To simplify the calculations, we restrict our discussion to orbital motion in the equatorial plane,  $\theta = \frac{\pi}{2}$. Orbits in other planes can be obtained by appropriately rotating the equatorial plane. In the equatorial case, there are $\Dot{\theta}=0$ and $L=L_z$. We are not concerned with the solution for the time coordinate $t$, and the orbital equations (\ref{Seom1}) - (\ref{SR}) further reduce to 
\begin{equation}\label{Seq1}
    \left( \frac{dr}{d\varphi} \right)^2=\frac{E^2-m^2}{L^2}r^4+2m^2\frac{M}{L^2}r^3-r^2+2M r.
\end{equation}
Since the condition $\left( \frac{dr}{d\varphi} \right)^2 \geq 0$ must hold, particles arriving from infinity must satisfy $E \geq m$.
The case $E = m$ corresonds to particles falling freely from rest and is not a critical orbit. Therefore, we restrict our analysis to $E > m$. Defining $u = 1/r$, Eq. (\ref{Seq1}) can be rewritten as
\begin{equation}\label{SchTime}
    \left(\frac{du}{d\varphi}\right)^2=2M u^3-u^2+\frac{2m^2M}{L^2}u+\frac{E^2-m^2}{L^2}\equiv f(u).
\end{equation}
\begin{figure*}
\centering
  \includegraphics[width=.55\textwidth]{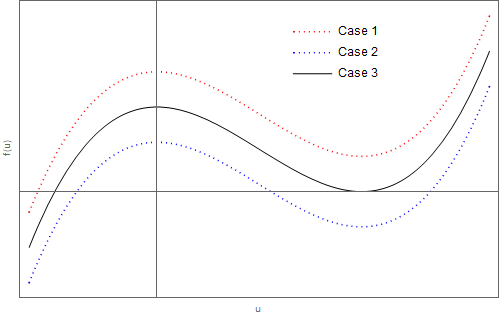}
\caption{We consider all possible root configurations of $f(u)=0$. In Case 1, where there are no real roots for $u>0$, a photon incident from infinity is absorbed by the black hole. In Case 2, characterized by two distinct real roots for $u>0$, the photon is scattered. Case 3 corresponds to the critical orbit, where a double root occurs at $u_c>0$. In this critical case, as we will discuss later, the photon is neither absorbed nor scattered but instead approaches a circular orbit at a finite radius $r_c=\frac{1}{u_c}$.}
\label{Fig:fu}
\end{figure*}
\subsection{null geodesic}
In the case of a null geodesic $(m=0)$, the function $f(u)$ simplifies to:
\begin{equation}\label{nullfu1}
    f(u)=2Mu^3-u^2+\frac{E^2}{L^2}.
\end{equation}
The orbital types are classified based on the roots of $f(u)=0$. 
Since $f(0)= \frac{E^2}{L^2}>0$, there is always a negative root. The three possible orbital types — absorption, scattering, and critical — correspond to the nature of the remaining roots; see Fig. \ref{Fig:fu}. For the critical orbit, the remaining two positive roots coincide at $u_c$, that is:
\begin{equation}\label{CriCond}
    f(u_c)=0,\quad f'(u_c)=0,
\end{equation}
where prime denotes differentiation with respect to the variable. Solving these equations yields
\begin{equation}
    u_c=\frac{1}{3M},\quad \frac{E_c^2}{L_c^2}=\frac{1}{27M^2}.
\end{equation}
The critical radius $r_c=\frac{1}{u_c}=3M$ defines the photon sphere, whereas the critical impact parameter $R_c=\frac{L_c}{E_c}=3\sqrt{3}M$ corresponds to the radius of the black hole's shadow as seen by an observer at infinity.

\begin{figure*}
\centering
  \includegraphics[width=.45\textwidth]{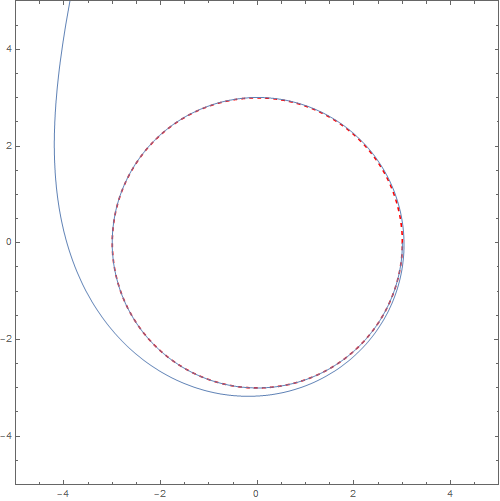}
\caption{Critical orbits of null geodesics in the Schwarzschild metric are shown. The mass is set to $M=1$, with the constant $\varphi_0=0$. The dashed line indicates the location of the photon sphere.}
\label{Fig:SchNull}
\end{figure*}

Substituting the critical impact parameter into (\ref{nullfu1}), the orbital equation reduces to 
\begin{equation}
    \left(\frac{du}{d\varphi}\right)^2=2M(u-u_c)^2(u+\frac{u_c}{2}),
\end{equation}
where the negative root is $u_{*}=-\frac{u_c}{2}$. The solution is given by: 
\begin{equation}
    u=-\frac{1}{6M}+\frac{1}{2M}\tanh^2\frac{1}{2}(\varphi-\varphi_0),
\end{equation}
where $\varphi_0$ is an integration constant. The critical null geodesic is shown in Fig. \ref{Fig:SchNull}, depicting the trajectory of a massless particle originating at infinity and spiraling inward, asymptotically approuching the circular at $r_c=3M$. 

\subsection{timelike geodesic}
\begin{figure*}
\centering
  \includegraphics[width=.45\textwidth]{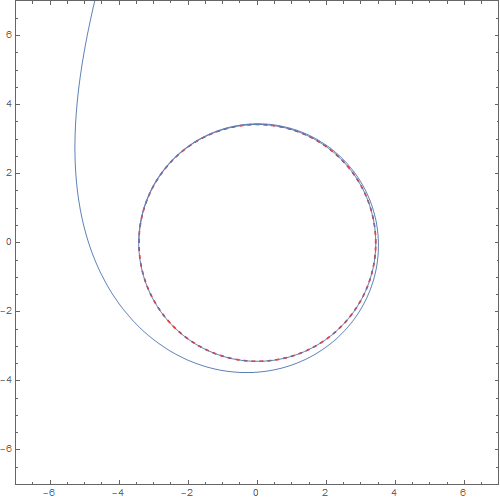}
\caption{Critical orbits for timelike geodesics in the Schwarzschild metric. The parameters are chosen as $M=1,\varphi_0=0$ and $u_c=\frac{7}{24}$. The dashed line marks the critical radius $r_c=\frac{1}{u_c}$.}
\label{Fig:SchTime}
\end{figure*}

Unlike the case of null geodesics, the critical radius for timelike geodesics depends on the parameters $E$ and $L$.\footnote{When a triple root occurs, the critical radius is uniquely given by $u_{c}=\frac{1}{6M}$. However, this corresponds to the bound case with $E^2<m^2$. } When $E^2>m^2$, we have $f(0)=\frac{E^2-m^2}{L^2}>0$, which imples the existence of a single negative root. If the other two real roots coincide, then the conditions (\ref{CriCond}) are also satisfied, namely
\begin{align}
    2M u_c^3-u_c^2+\frac{2m^2M}{L^2}u_c+\frac{E^2-m^2}{L^2}&=0,\\
    3M u_c^2-u_c+\frac{m^2M}{L^2}&=0.
\end{align}
The solutions are given by
\begin{align}
    L_c^2&=\frac{m^2M}{u_c-3Mu_c^2},\label{a1}\\
    E_c^2-m^2&=\frac{(4Mu_c-1)u_cm^2M}{1-3Mu_c}.\label{a2}
\end{align}
The conditions $L_c^2>0$ and $E_c^2-m^2>0$ imply $u_c\in(\frac{1}{4M},\frac{1}{3M})$. Substituting the critical impact parameter into equation  (\ref{SchTime}) reduces orbital equation to 
\begin{equation}
    \left(\frac{du}{d\varphi}\right)^2=2M(u-u_c)^2(u+2u_c-\frac{1}{2M}),
\end{equation}
where the negative root is $u_{*}=\frac{1}{2M}-2u_c$ and $u_c$ lies in the interval $(\frac{1}{4M},\frac{1}{3M})$. Mathematical analysis shows that
\begin{equation}
    u=u_c-\frac{1}{M}\frac{6Mu_c-1}{1+\cosh(\sqrt{6Mu_c-1}(\varphi-\varphi_0) )}.
\end{equation}
The critical timelike geodesic in the Schwarzschild spacetime is shown in Fig. \ref{Fig:SchTime}.

In the accretion model of Rioseco and Sarbach \cite{Rioseco2017-1}, the key observable quantities are the particle current density $J_{\mu}$ and the energy-momentum tensor $T_{\mu\nu}$, which are defined by
\begin{align}
J_{\mu}&=\int p_{\mu}f(x,p)\mathrm{dvol}_x(p),\label{Ober1}\\
T_{\mu\nu}&=\int p_{\mu}p_{\nu}f(x,p)\mathrm{dvol}_x(p),\label{Ober2}
\end{align}
where $f(x,p)$ is the distribution function and $\mathrm{dvol}_x(p)=\sqrt{-\det[g^{\mu\nu}]}dp_tdp_rdp_{\theta}dp_{\varphi}$ is the volume element. Introducing the action-angle momenta $(m,E,l_z,l)$, the volume element can be re-expressed as $\mathrm{dvol}_x(p)\propto dmdEdLdL_z $. For the integral over $dm$, it is generally assumed that the Vlasov gas consists of a large number of identical point particles with mass $m_0$, and therefore the distribution function includes a factor $\delta(m-m_0)$. For the integral over $dE$, unbound orbits statisfy $E>m_0$. For the integral over $dL_z$, the integration limits are $L_z\in[-L\sin\theta, L\sin\theta]$. The remaining integrals over $dL$ separate the observable quantities $J_{\mu}$ and $T_{\mu\nu}$ into three parts: the absorbed, the scattered, and the critical parts, which are:
\begin{align}
  J_{\mu}(u) &=J^{abs}_{\mu}+J^{scat}_{\mu}+J^{cri}_{\mu},  \\
  T_{\mu\nu}(u)&=T^{abs}_{\mu\nu}+T^{scat}_{\mu\nu}+T^{cri}_{\mu\nu}.
\end{align}
Since the critical part occurs at $L=L_c$, we have $\int_{L_c}^{L_c}f(l)dl=0$, so the critical part does not contribute substantially. We can re-express (\ref{a1}) and (\ref{a2}) as
\begin{equation}
    L_c^2=\frac{M^2 \left(27 E^4-36 E^2 m^2+8 m^4+ E  (9 E^2-8 m^2)^{3/2}\right)}{2 \left(E^2-m^2\right)}.
\end{equation}
For the absorption component, the interval is $L\in [0,L_c]$. For the scattering component, the interval is $L\in [L_c, L_{m}]$, where $L_{m}$ is chosen so that $f>0$ throughout. Solving $f=0$ yields
\begin{equation}
    L_m^2=\frac{2m^2Mu+E^2-m^2}{u^2-2M u^3}.
\end{equation}
Thus, we have obtained all integration intervals. In fact, we do not need to express them explicitly in the form $L_c(E)$, as the relation is already fully specified by equations (\ref{a1}) and (\ref{a2}). Therefore, we will not discuss the integration interval of the accretion model separately in the subsequent sections.

\section{Charged particles traveling in a Reissner-Nordstr\"{o}m space-time}
In this section, we study the motion of charged particles in the Reissner-Nordstr\"{o}m black hole spacetime with $a=0$. The equations (\ref{eom1}) - (\ref{R}) governing this motion reduce to 
\begin{align}
  \Dot{t}&=\frac{r}{\Delta}(Er-eQ),\\
  \Dot{\varphi}&=\frac{L_z}{r^2\sin^2\theta},\label{Reom1}\\
  \Dot{\theta}^2&=\frac{1}{r^4}\left(L^2-\frac{L_z^2}{\sin^2\theta} \right),\\
  r^4\Dot{r}^2&=(E^2-m^2)r^4+2(m^2M-eQE) r^3-L^2\Delta.\label{RR}
\end{align}
As in Schwarzschild spacetime, the square of the angular momentum $L^2$ is conserved. Therefore, the worldlines are confined to a single plane. Note that a charged test particle in Reissner-Nordstr\"{o}m spacetime does not, in general, follow geodesics. The orbital equation in the plane is given by $\left(\frac{du}{d\varphi}\right)^2=f(u)$, where
\begin{equation}
      f(u)=-Q^2u^4+2M u^3-\left(1+\frac{m^2-e^2}{L^2}Q^2\right)u^2+2\frac{m^2M-eQE}{L^2}u+\frac{E^2-m^2}{L^2}.
\end{equation}
Now, $f(u)$ is a quartic function in $u$. Given that $f(0) > 0$, the equation $f(u) = 0$ must have one negative real root $u_{*}^{-}$ and one positive real root $u_{*}^{+}$. Therefore, the zeros of the function $f(u)$ may fall into one of the following cases: (i) four real roots, three of which coincide; (ii) four real roots, two of which coincide; (iii) two real roots and two complex roots. This paper primarily concerns with cases (i) and (ii). Both cases satisfy $f(u_c)=0,f'(u_c)=0$.
\subsection{null geodesic}
The null geodesic $(m=e=0)$ in the Reissner-Nordstr\"{o}m geometry is governed by the function
\begin{equation}
 f(u) =-Q^2u^4+2M u^3-u^2+\frac{E^2}{L^2}.  
\end{equation}
In the critical case $f(u_c)=0$ and $f'(u_c)=0$, it is straightforward to obtain
\begin{align}
   u_c&=\frac{3M\pm \sqrt{9M^2-8Q^2}}{4Q^2},\\
   \frac{E_c^2}{L_c^2}&=-\frac{27 M^4-36 M^2 Q^2+8 Q^4\pm9 M^3 \sqrt{9 M^2-8 Q^2}\mp8 M Q^2 \sqrt{9 M^2-8 Q^2}}{32 Q^6}.
\end{align}
Photons incident from infinity will first encounter the larger critical radius $r_c=\frac{1}{u_c}$; this is the case we focus on in the subsequent analysis. 

As in the Schwarzschild case, the shadow radius of a Reissner–Nordstr\"{o}m black hole, as seen by a distant observer, is given by 
$R_c=\frac{L_c}{E_c}$. For small values of the charge $Q$, a Taylor series expansion yields
\begin{align}
u_c &= \frac{1}{3M} \left( 1 + \frac{2 Q^2}{9 M^2} + O\left(\frac{Q^4}{M^4}\right) \right), \\
R_c &= 3\sqrt{3} M \left( 1 - \frac{Q^2}{6 M^2} + O\left(\frac{Q^4}{M^4}\right) \right).
\end{align}
Thus, relative to an uncharged black hole, a charged black hole has a smaller photon sphere and therefore casts a smaller shadow. In the extremal case $Q=M$, the photon sphere and shadow attain their minimal radii
\begin{equation}
r_c^{\text{ex}} = \frac{1}{u_c^{\text{ex}}} = 2M, \quad R_c^{\text{ex}} = 4M.
\end{equation}

Given the condition for the existence of a horizon $Q < M$, the case of three coincident real roots cannot occur. Thus, in the critical case, the function $f(u)$ can only degenerate into
\begin{equation}\label{RNnull}
    f(u)=-Q^2(u-u_c)^2(u-u_{*}^{-})(u-u_{*}^{+}),
\end{equation}
where 
\begin{align}
   u_{*}^{\pm}&=\frac{M+\sqrt{9M^2-8Q^2}\pm2\sqrt{M(M+\sqrt{9M^2-8Q^2})}}{4Q^2},
\end{align}
and $u_{*}^{-}<0<u_c<u_{*}^{+}$. Solving the equation $\frac{du}{d\varphi}=\sqrt{f(u)}$, we obtain
\begin{equation}\label{RNdoub}
    u=u_c-\frac{2 (u_{*}^{-}-u_c) (u_c-u_{*}^{+})}{u_{*}^{-}+u_{*}^{+}-2 u_c+(u_{*}^{+}-u_{*}^{-}) \cosh \left(Q (\varphi -\varphi_0) \sqrt{(u_c-u_{*}^{-}) (u_{*}^{+}-u_c)}\right)}.
\end{equation}
The plot of this solution closely resembles Fig. \ref{Fig:SchNull} and is therefore omitted here.

\subsection{timelike worldline}
The conditions for timelike critical orbits are as follow:
\begin{align}
    -Q^2u^4+2Mu^3-\left(1+\frac{m^2-e^2}{L^2}Q^2 \right)u^2+2\frac{m^2M-eQE}{L^2}u+\frac{E^2-m^2}{L^2}&=0,\\
    -2Q^2u^3+3Mu^2-\left(1+\frac{m^2-e^2}{L^2}Q^2 \right)u+\frac{m^2M-eQE}{L^2}&=0.
\end{align}
If the critical orbits have radius $r_c=\frac{1}{u_c}$, then the energy $E$ and the angular momentum are given by
\begin{align}
   E_c&=\frac{1}{2Y_c}\left(\Delta_c\sqrt{4m^2Y_c+e^2Q^2u_c^2}+eQu_cP_c \right),\\
   L_c^2&=\frac{1}{2u_cY_c^2}\left(2m^2(M-Q^2u_c)Y_c-eQ\Delta_c(\sqrt{4m^2Y_c+e^2Q^2u_c^2}-eQu_c)  \right),\label{Lceq}
\end{align}
where
\begin{align}
    \Delta_c&=1-2Mu_c+Q^2u_c^2,\\
    Y_c&=1-3Mu_c+2Q^2u_c^2,\\
    P_c&=1-4Mu_c+3Q^2u_c^2.
\end{align}
These equations require that $4m^2Y_c+e^2Q^2u_c^2>0$, which implies $u_{c}^{-}<u_c<u_{c}^{+}$, where
\begin{equation}
   u_c^{\pm}=\frac{2 \left(3 m^2 M\pm\sqrt{-e^2 m^2 Q^2+9 m^4 M^2-8 m^4 Q^2}\right)}{e^2 Q^2+8 m^2 Q^2}. 
\end{equation}

\begin{figure*}
\centering
  \includegraphics[width=.45\textwidth]{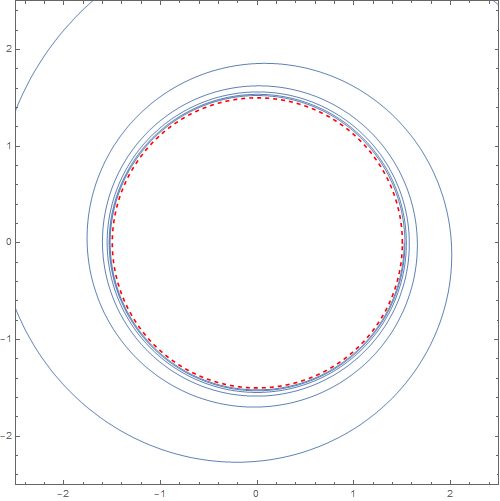}
\caption{Shown are the critical orbits for case (i) of a timelike worldline in the Reissner–Nordstr\"{o}m metric. The parameters are chosen as $u_c=\frac{2}{3M},e=1.5m,Q=\frac{3 m M}{\sqrt{e^2+8 m^2}},M=1,\varphi_0=0$. The dashed line indicates the location of the critical radius $r_c=\frac{1}{u_c}$.}
\label{Fig:RNTimeTrip}
\end{figure*}

 A fundamentally new scenario arises—the case of triply coincident roots. For this to occur, the additional constraint $f''(u)=0$ must be satisfied, namely,
 \begin{equation}
     6Q^2u^2-6Mu+1+\frac{m^2-e^2}{L^2}Q^2=0.
 \end{equation}
Eliminating $L^2$ using Eq. (\ref{Lceq}), we obtain: 
\begin{align}
   (6Q^2 u_c^2&-6M u_c+1)[2m^2(M-Q^2u_c)Y_c-eQ\Delta_c(\sqrt{4m^2Y_c+e^2Q^2u_c^2}-eQu_c) ]\nonumber\\
   &+2(m^2-e^2)Q^2u_cY_c^2=0.    
\end{align}
Note that, while the above equation may admit multiple solutions, only those with $u_c$ satisfying $u_{c}^{-}<u_c<u_{c}^{+}$ and corresponding to a triple root are valid. In particular, when $e=0$, the above equation reduces to
\begin{equation}
    4Q^4u_c^3-9MQ^2u_c^2+6M^2u_c-M=0.
\end{equation}
This equation determines the minimum radius of a stable circular orbit in Reissner-Nordstr\"{o}m geometry. In the critical case (i), the planar orbital equation takes the form:
\begin{equation}
    \left(\frac{du}{d\varphi}\right)^2=-Q^2(u-u_c)^3(u-\frac{2M-3Q^2u_c}{Q^2}),
\end{equation}
where the negative root $u_{*}^{-}=\frac{2M-3Q^2u_c}{Q^2}$ shows that the triple root satisfies $u_c>\frac{2M}{3Q^2}$. The solution is 
\begin{equation}
    u=u_c+\frac{2(M-2Q^2u_c)}{Q^2+(M-2Q^2u_c)^2(\varphi-\varphi_0)^2}.
\end{equation}
An example of this orbit is plotted in Fig. \ref{Fig:RNTimeTrip}.

In the critical case (ii), the planar orbital equation takes the following form:
\begin{equation}
     \left(\frac{du}{d\varphi}\right)^2=-Q^2(u-u_c)^2(u-u_{*}^{-})(u-u_{*}^{+}),
\end{equation}
where
\begin{equation}
   u_{*}^{\pm}=\frac{L_c^2(M-Q^2u_c)\pm|L_c|\sqrt{Q^4 \left(m^2-e^2-2 L_c^2 u_c^2\right)+L_c^2 Q^2 (2 M u_c+1)+L_c^2 M^2 }}{ L_c^2Q^2}, 
\end{equation}
where $L_c^2$ is given by Eq. (\ref{Lceq}). This equation has the same form as the null-geodesic orbital equation (\ref{RNnull}); hence, its solution can be written as Eq. (\ref{RNdoub}), but with different parameters $u_c$ and $u_{*}^{\pm}$.

\section{Neutral particles traveling in a Kerr space-time}
In this section, we study the critical orbits of neutral particles in the Kerr geometry. For $Q=e=0$, equations (\ref{eom1}) - (\ref{R}) reduce to 
\begin{align}
    \rho^2\Dot{t}&=\frac{1}{\Delta}\left[E((r^2+a^2)^2-\Delta a^2\sin^2\theta)-2ML_z ar) \right],\label{eomK1}\\
    \rho^2\Dot{\varphi}&=\frac{1}{\Delta}\left[2Ma Er+ \frac{L_z}{\sin^2\theta}(\Delta-a^2\sin^2\theta)\right],\\
    \rho^4\Dot{\theta}^2&\equiv \Theta=D+\cos^2\theta \big(a^2(E^2-m^2)-\frac{L_z^2}{\sin^2\theta} \big),\\
     \rho^4\Dot{r}^2&\equiv R=(E^2-m^2)r^4+2m^2Mr^3+[a^2(E^2-m^2)-L_z^2-D]r^2\nonumber\\
     &+2M((L_z-aE)^2+D)r-a^2D.\label{eomK2}
\end{align}
Here, Eq. (\ref{R}) reduces to Eq. (\ref{eomK2}).

In the following discussion, we introduce new variables:
 \begin{equation}
   \xi=\frac{L_z}{\sqrt{E^2-m^2}},\quad \eta=\frac{D}{E^2-m^2},\quad \chi^2=\frac{E^2}{E^2-m^2}. 
 \end{equation}
Then, the functions $R$ and $\Theta$ can be rewritten as follow:
\begin{align}
    \frac{R}{E^2-m^2}&=r^4+2M(\chi^2-1)r^3+(a^2-\xi^2-\eta)r^2+2M[\eta+(\xi-a\chi)^2]r-a^2\eta,\\
    \frac{\Theta}{E^2-m^2}&=\eta+a^2\cos^2\theta-\xi^2\cot^2\theta.\label{thetaK}
\end{align}
We begin by considering the solution in the $\theta$-direction, before discussing the $r$-motion. The $\theta$-motion is governed by the integral $\int\frac{d\theta}{\sqrt{\Theta}}$. For this integral to be real-valued, the condition $\eta+(a-\xi)^2 \geq 0$ must be satisfied. Let $\cos\theta=\mu$ be the variable of integration; then $\int\frac{d\theta}{\sqrt{\Theta}}=-\int\frac{d\mu}{\sqrt{\tilde{\Theta}}}$ where
\begin{equation}
    \frac{\tilde{\Theta}}{E^2-m^2}=\eta-(\xi^2+\eta-a^2)\mu^2-a^2\mu^4.\label{SolKTheta}
\end{equation}
The behavior of the $\theta$-motion is categorized into three cases: $\eta>0, \eta<0$, and $\eta=0$. 

\begin{enumerate}
    \item $\eta>0$: We can re-express $\tilde{\Theta}$ as follow:
\begin{align}
  \frac{\tilde{\Theta}}{E^2-m^2}&=a^2(\mu_1^2+\mu^2)(\mu_2^2-\mu^2),
\end{align}
where 
\begin{align}
    \mu_1^2&=\frac{1}{2a^2}\left(\sqrt{(\xi^2+\eta-a^2)^2+4a^2\eta_c}+(\xi^2+\eta-a^2)  \right),\label{mu1}\\
       \mu_2^2&=\frac{1}{2a^2}\left(\sqrt{(\xi^2+\eta-a^2)^2+4a^2\eta}-(\xi^2+\eta-a^2)  \right).\label{mu2}
\end{align}
The requirement $\tilde{\Theta}\geq 0$ implies that $0\leq \mu^2\leq \mu_2^2$. Here, $\mu^2=0$ corresponds to the equatorial plane $\theta=\frac{\pi}{2}$. When $\mu^2=\mu_2^2$, the orbits attain the turning angles at $\theta=\frac{\pi}{2}\pm \theta_m$, where $\theta_m=\arccos \mu_2$. In other words, these orbits cross the equatorial plane and oscillate symmetrically about it. The integration yields
 \begin{align}
\frac{1}{\sqrt{E^2-m^2}}\int^{\mu}\frac{d\mu}{\sqrt{\tilde{\Theta}}}&=\frac{1}{a \sqrt{\mu_1^2+\mu_2^2}}\text{EllipticF}\left[\arcsin\frac{\sqrt{\mu_2^2-\mu^2}}{\mu_2},\frac{\mu_2^2}{\mu_1^2+\mu_2^2} \right]+C,\label{ThetaInt}
\end{align}
where $\text{EllipticF}[\phi,m]$ denotes the incomplete elliptic integral of the first kind, and $C$ is an integration constant.

\item $\eta=0$: In this special case, we have
\begin{equation}
  \frac{\tilde{\Theta}}{E^2-m^2}=a^2\mu^2(\mu_m^2-\mu^2),
\end{equation}
where $\mu_m^2=1-\frac{\xi^2}{a^2}$, and $0\leq \mu^2\leq\mu_m^2\leq 1$. The integration yields:
 \begin{align}
\frac{1}{\sqrt{E^2-m^2}}\int^{\mu}\frac{d\mu}{\sqrt{\tilde{\Theta}}}&=-\frac{1}{a\mu_m}\arctanh\sqrt{1-\frac{\mu^2}{\mu_m^2}}+C.
\end{align}

    \item $\eta<0$: In this case, the function $\tilde{\Theta}$ can be re-expressed as
\begin{equation}
   \frac{\tilde{\Theta}}{E^2-m^2}=a^2(\mu_2^2-\mu^2)(\mu^2-\mu_1^2),  
\end{equation}
where
\begin{equation}
    \mu_{1,2}=\frac{1}{2a^2}\left( (|\eta|+a^2-\xi^2)\pm\sqrt{(|\eta|+a^2-\xi^2)^2-4a^2|\eta|}   \right).
\end{equation}
Thus, the parameter $\mu^2$ is restricted to $0\leq\mu_1^2\leq\mu^2\leq \mu_2^2\leq 1$. In fact, if the $\eta<0$ case actually happens, it requires very stringent conditions: $|\eta|<a^2$ and $0<|\xi|\leq|a|$. In such cases, the test particle will inevitably enter the region of negative $r$, and its worldline will terminate this negative $r$ region. However, to date our analysis has not identified a physically feasible set of parameters that would realize this scenario. We include it here solely as a mathematical possibility for completeness. The integration yields
 \begin{align}
\frac{1}{\sqrt{E^2-m^2}}\int^{\mu}\frac{d\mu}{\sqrt{\tilde{\Theta}}}&=\frac{1}{a\mu_2}\text{EllipticF}\left[\arcsin \sqrt{\frac{\mu_2^2(\mu^2-\mu_1^2)}{\mu^2(\mu_2^2-\mu_1^2)}},\frac{\mu_2^2-\mu_1^2}{\mu_2^2}\right]+C.
\end{align}    
\end{enumerate}

\subsection{null geodesic}
We then consider the radial motion. For massless particles  traveling in Kerr geometry ($m=0$ and $\chi=1$), $R$ reduces to 
\begin{align}
    \frac{R}{E^2}&=r^4+(a^2-\xi^2-\eta)r^2+2M[\eta+(\xi-a)^2]r-a^2\eta.\label{RK}
\end{align}
We consider the critical orbits $R=0$ and $R'=0$
\begin{align}
    r^4+(a^2-\xi^2-\eta)r^2+2M[\eta+(\xi-a)^2]r-a^2\eta&=0,\label{eqK1}\\
    4r^3+2(a^2-\xi^2-\eta)r+2M[\eta+(\xi-a)^2]&=0.\label{eqK2}
\end{align}
The equations above yield two distinct pairs of solutions for $\xi$ and $\eta$; one is given below
\begin{equation}\label{solk1}
\xi_c = \frac{a^2 + r_c^2}{a}, \quad \eta_c = -\frac{r_c^4}{a^2};
\end{equation}
and the other is
\begin{equation}\label{solk2}
\xi_c = \frac{r_c^2(r_c - 3M) + a^2(M + r_c)}{a(M - r_c)}, \quad \eta_c = \frac{r_c^3\left(   4a^2M- (r_c - 3M)^2 r_c\right)}{(r_c - M)^2 a^2}.
\end{equation}
Substituting solution~(\ref{solk1}) into equation~(\ref{thetaK}), we find that $\Theta = -\frac{\rho^2 E^2}{a^2 \sin^2\theta} \leq 0$. Since physical motion requires $\Theta \geq 0$, this condition can only be satisfied if $\theta=\theta_0$ remains constant, corresponding to shear-free null congruences. Our primary interest lies in the second set of solutions (\ref{solk2}).

\begin{figure*}
\centering
  \includegraphics[width=.45\textwidth]{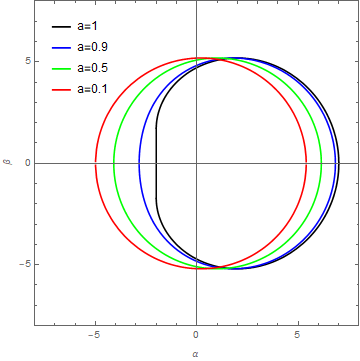}
\caption{The shadow of a Kerr black hole, as observed by a distant observer in the equatorial plane, is shown. The mass of the black hole is set to $M=1$.}
\label{Fig:KerrShad}
\end{figure*}

Celestial coordinates were introduced in \cite{Bardeen1973}
\begin{align}
  \alpha&=\lim_{r_0\rightarrow \infty}\left(-r_0^2\sin\theta_0\frac{d\varphi}{dr}\right)=-\xi \csc\theta_0,\\
   \beta&=\lim_{r_0\rightarrow \infty} \left(r_0^2\frac{d\theta}{dr}\right)=\pm (\eta+a^2\cos^2\theta_0-\xi^2\cot^2\theta_0)^{1/2}.
\end{align}
Here, $r_0$ is the distance from the black hole to the observer, and $\theta_0$ is the observer's viewing angle. If the observer lies in the equatorial plane, $\theta_0=\frac{\pi}{2}$, the celestial coordinates can be written as $\alpha=-\xi$ and $\beta=\pm\sqrt{\eta}$. The apparent shape of the black hole is determined by the critical coordinates $(\alpha_c,\beta_c)$. Figure \ref{Fig:KerrShad} depicts the shadow of the Kerr black hole as viewed from the equatorial plane.

\begin{figure*}
\centering
  \includegraphics[width=.45\textwidth]{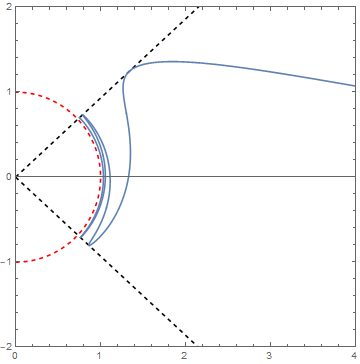}
  \includegraphics[width=.45\textwidth]{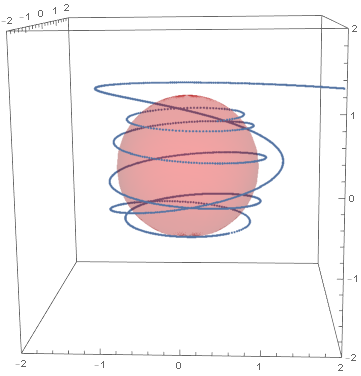}
\caption{Critical null orbits with triple coincident roots in Kerr spacetime. The left panel illustrates the trajectory of a null geodesic in the $(r,\theta)$ plane; the right panel shows the trajectory in three-dimensional space. The mass of the black hole is set to $M=1$, and the initial conditions are $r_{i}=10,\mu_{i}=0$, and $\varphi_i=0$. The red dashed line (ball) denotes the phonton sphere, and the black dashed lines correspond to $\mu=\pm \mu_{m}$.}
\label{Fig:KerrNull1}
\end{figure*}

There are two types of critical orbits, defined by the root structures of $R=0$: (i) triple coincident roots and (ii) double coincident roots. In addition to Eqs. (\ref{eqK1}) and (\ref{eqK2}), the case of triple coincident roots requires that the condition $R''=0$ also be satisfied, i.e.,
\begin{equation}
 6r^2+a^2-\xi^2-\eta=0.   
\end{equation}
Define $\mu_0^2=\frac{M r_c}{3M^2-3M r_c+r_c^2}$; one can then obtain the solution
\begin{align}
    \xi^{*}_c&=\frac{3M^2-r_c^2}{M}\mu_0,\quad
    \eta^{*}_c=3r_c^2\mu_0^2,\quad
    a^{*}_c=\frac{r_c}{\mu_0}.\label{KTriple}
\end{align}
Rewriting relation (\ref{KTriple}) yields
\begin{equation}
\frac{a^{*}_c}{M}=\sqrt{\frac{r_c}{M}}\sqrt{3-3\frac{r_c}{M}+\left(\frac{r_c}{M}\right)^2}.
\end{equation} 
This expression reveals a monotonic relationship: $a^{*}_c$  increases with $r_c$. Given the constraints for the existence of the horizon, $0\leq a\leq M$, and that the photon sphere lies outside the horizon, $r_{+}\leq r_c$, it follows that the case of triply coincident roots occurs only when $a^{*}_c=M$ and $r_c=r_{+}=M$, which implies $\mu_0^2=1$, with $\xi_c=2M, \eta_c=3M^2$. Substituting these solutions into $R$ and $\tilde{\Theta}$ yields
\begin{align}
   \frac{R}{E^2}&= (r-M)^3(r+3M),\\
 \frac{\tilde{\Theta}}{E^2}&=M^2(2\sqrt{3}+3+\mu^2)(2\sqrt{3}-3-\mu^2).\label{Rrd}
\end{align}
Mathematically, we obtain
\begin{align}
   \int \frac{dr}{E\sqrt{R}}&=- \frac{\sqrt{(r-M)^3(r+3M)}}{2M(r-M)^2}+C_1,\label{Rint1}\\
   \int \frac{d\mu}{E\sqrt{\tilde{\Theta}}}&=\frac{1}{M\sqrt{2\sqrt{3}+3}}
\text{EllipticF}\left[\arcsin\frac{\mu}{\sqrt{2\sqrt{3}-3}},-\frac{2\sqrt{3}-3}{2\sqrt{3}+3}\right]+C_2,
\end{align}
where $C_1, C_2$ are integration constants. Since $\eta=3M^2>0$, the $\theta$-motion oscillates between $(\frac{\pi}{2}-\theta_{m},\frac{\pi}{2}+\theta_{m})$, where $\theta_{m}=\arccos \mu_{m}$. The constraint on $(r,\theta)$ given in (\ref{constraint}) can now be integrated to yield
\begin{equation}
\int_{r_i}^r \frac{dr}{\sqrt{R}}= \left(\int_{\mu_i}^{\mu_m^1}-\int_{\mu_m^1}^{\mu_m^2}...+(-1)^{n-1}\int_{\mu_m^n}^{\mu}  \right)\frac{d\mu}{\sqrt{\tilde{\Theta}}} ,
\end{equation}
where $r_i, \mu_i$ are the initial conditions. Here, the limits $\mu_m^{n}$ alternate between $\mu_m$ and $-\mu_m$, namely $\mu_m^{2k+1}=\pm\mu_m$ and $\mu_m^{2k}=\mp\mu_m$. The number of integrals $n$ on the right-hand side equals the number of oscillations, which in turn depends on the final radial coordinate $r$. The solution for $\varphi$ can be written as
\begin{align}
 \varphi&=\frac{(r-3M)}{6}\sqrt{\frac{r+3M}{(r-M)^3}}\nonumber\\
 &-\frac{2}{\sqrt{2\sqrt{3}+3}}\text{ElliptiPi}\left[2\sqrt{3}-3,\arcsin{\frac{\mu}{\sqrt{2\sqrt{3}-3}}},-\frac{2\sqrt{3}-3}{2\sqrt{3}+3} \right]+C,
\end{align}
where $\text{ElliptiPi}[n,\varphi,m]$ denotes the incomplete elliptic integral of the third kind. The numerical solution is shown in Fig. \ref{Fig:KerrNull1}.

\begin{figure*}
\centering
\setlength{\tabcolsep}{20pt} 
  \renewcommand{\arraystretch}{1.5} 
  \begin{tabular}{cc} 
    \includegraphics[width=0.35\textwidth]{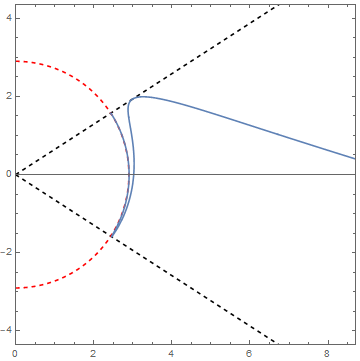} &
    \includegraphics[width=0.35\textwidth]{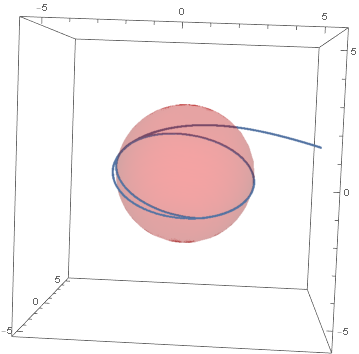} \\    
    \includegraphics[width=0.35\textwidth]{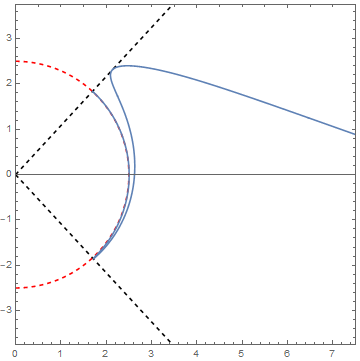} &
    \includegraphics[width=0.35\textwidth]{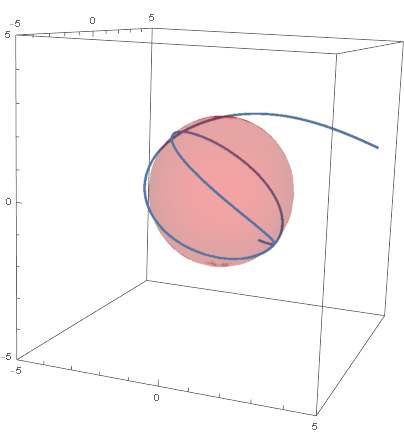} \\    
    \includegraphics[width=0.35\textwidth]{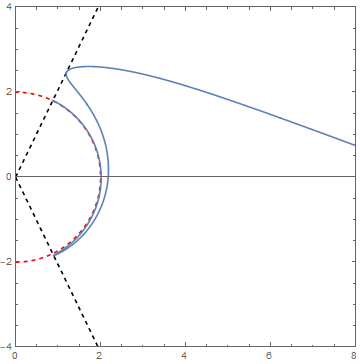} &
    \includegraphics[width=0.35\textwidth]{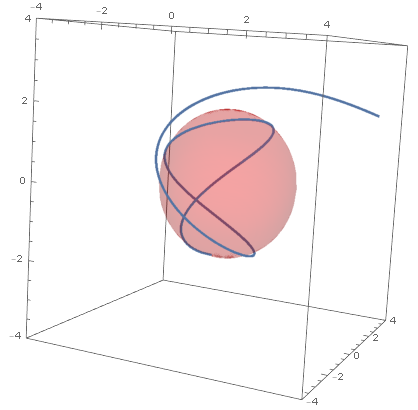} 
  \end{tabular}
\caption{Critical null orbits corresponding to double-coincident root in Kerr spacetime. The left diagram illustrates the trajectory of a null geodesic in the $(r,\theta)$ plane, while the right diagram shows the trajectory of a null geodesic in three-dimensional space. The mass of the black hole is set to $M=1$, and the initial conditions are $r_{i}=10,\mu_{i}=0,\varphi_i=0$. The parameters for each row of images (from top to bottom) are $a=0.1,r_c=2.9; a=0.5,r_c=2.5;$ and $a=0.9,r_c=2.0$. The red dashed line (ball) is located at $r=r_c$, and the black dashed lines are at $\mu=\pm \mu_{m}$.}
\label{Fig:KerrNull2}
\end{figure*}

The critical null geodesics with double coincident roots can be analyzed in a similar manner. Substituting $\xi=\xi_c, \eta=\eta_c$ into Eqs. (\ref{RK}), we can rewrite $R$ as 
\begin{align}
     \frac{R}{E^2}&=(r-r_c)^2(r^2+2r_c r-\frac{a^2\eta_c}{r_c^2}).
\end{align}
Define $x=\frac{1}{r-r_c}$, we obtain
\begin{align}
\int^r\frac{dr}{E\sqrt{R}}&=\frac{1}{\sqrt{c}}\ln[\sqrt{c}\sqrt{cx^2+4r_cx+1}+cx+2r_c   ]  +C,\label{Rint2}
\end{align}
where $c=3r_c^2-\frac{a^2\eta_c}{r_c^2}$. The dependence between the coordinates $r$ and $\theta$ is given by Eq. (\ref{constraint}), and a detailed analysis of the $\theta$ motion is provided in the previous subsection. The solution for $\varphi$ is obtained by numerical integration. Critical null geodesics with double root are shown in Fig. \ref{Fig:KerrNull2}.  

\subsection{Timelike geodesic}
\begin{figure*}
\centering
\setlength{\tabcolsep}{20pt} 
  \renewcommand{\arraystretch}{1.5} 
  \begin{tabular}{cc} 
    \includegraphics[width=0.35\textwidth]{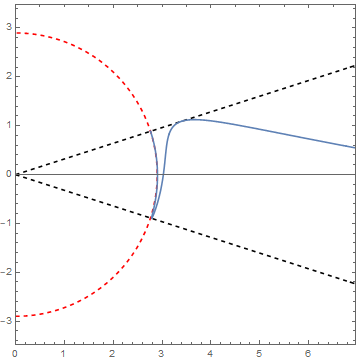} &
    \includegraphics[width=0.35\textwidth]{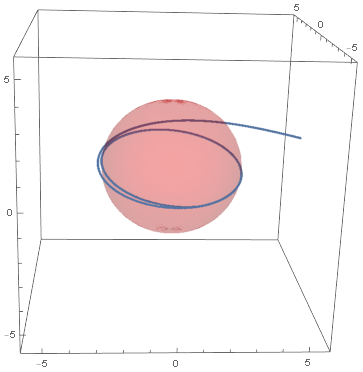} \\   
    \includegraphics[width=0.35\textwidth]{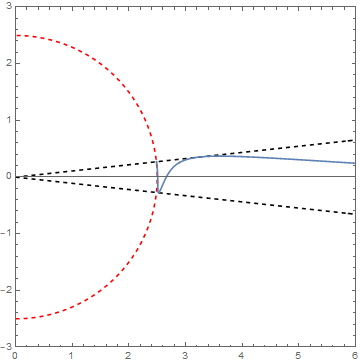} &
    \includegraphics[width=0.35\textwidth]{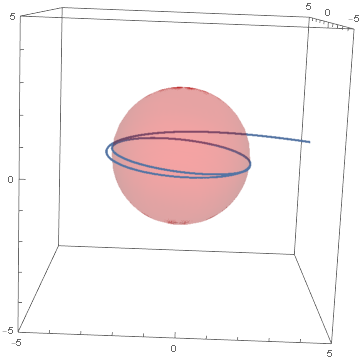} \\   
    \includegraphics[width=0.35\textwidth]{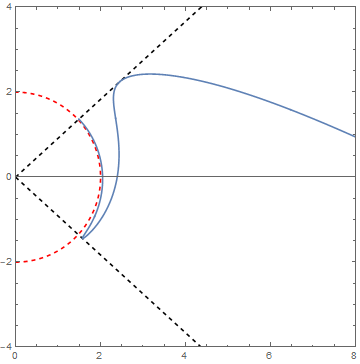} &
    \includegraphics[width=0.35\textwidth]{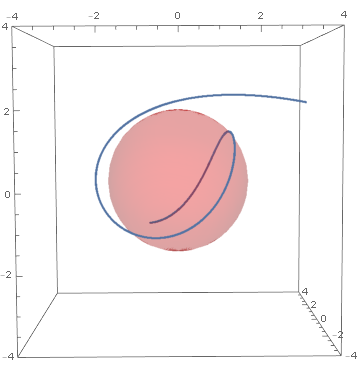} 
  \end{tabular}
\caption{Critical timelike geodesics in Kerr spacetime. The left diagram illustrates the trajectory of a timelike geodesic in the $(r,\theta)$ plane, while the right diagram shows the trajectory in three-dimensional space. The mass of the black hole is set to $M=1$, and the initial conditions are $r_{i}=10,\mu_{i}=0,\varphi_i=0$. The parameters for each row of images (from top to bottom) are $a=0.1,r_c=2.9,\chi=1.02; a=0.5,r_c=2.5,\chi=1.4; a=0.9,r_c=2.0,\chi=4$. The red dashed line (ball) is at $r=r_c$, and the black dashed line denotes $\mu=\pm \mu_{m}$.}
\label{Fig:KerrTimeLike}
\end{figure*}

For timelike geodesics with $m\neq 0$, we show 
in Appendix A that no triple-root unbound orbits exist when $r_c > r_+$.\footnote{Similar to trajectories in a Schwarzschild background, the triple-root scenario in Kerr spacetime occurs for bound orbits with $E^2<m^2$; see also \cite{Geoffrey2022}.} The conditions for a double root, $R=0$ and $R'=0$, yield
\begin{align}
    \eta_c&=-\xi_c^2-\frac{2aM\chi}{r_c-M}\xi_c-\frac{a^2 M \chi ^2+a^2 r_c+3 M r_c^2 \chi ^2-3 M r_c^2+2 r_c^3}{r_c-M},\\
    \xi_c&=\frac{ aM(r^2-a^2)\chi\pm\sqrt{a^2M^2(r_c^2-a^2)^2\chi^2+4K(r_c-M)a^2}}{a^2(r_c-M)},\\
    K&=r_c\Delta^2(r_c)+M\chi^2(r_c^4-4Mr_c^3+2a^2r_c^2+a^4 ).
\end{align}
The function $R$ can then be expressed as
\begin{align}
     \frac{R}{E^2-m^2}&=(r-r_c)^2\left(r^2+2(r_c+M(\chi^2-1)) r-\frac{a^2\eta_c}{r_c^2}\right).
\end{align}
Defining $x=\frac{1}{r-r_c}$, one can evaluate the integrals to obtain
\begin{equation}
    \int^r\frac{dr}{\sqrt{E^2-m^2}\sqrt{R}}=\frac{1}{\sqrt{c}}\ln[\sqrt{c}\sqrt{cx^2+2bx+1}+cx+b ]+C, \label{RintKN1}
\end{equation}
where
\begin{align}
    b&=2r_c+M(1-\chi^2),\\
    c&=3r_c^2+2Mr_c(\chi^2-1)-\frac{a^2\eta_c}{r_c^2}.
\end{align}
The numerical results are plotted in Fig. \ref{Fig:KerrTimeLike}.

\section{Charged particles traveling in a Kerr-Newman space-time}
In this section, we investigate the critical orbits of charged particles moving in a Kerr-Newman background. We define
 \begin{equation}
   \xi=\frac{L_z}{\sqrt{E^2-m^2}},\quad \eta=\frac{D}{E^2-m^2},\quad \chi^2=\frac{E^2}{E^2-m^2},\quad \tilde{e}=\frac{e}{\sqrt{E^2-m^2}}, 
 \end{equation}
and rewritten $R$ as
\begin{align}
    \frac{R}{E^2-m^2}&=r^4+2\left(M(\chi^2-1)-\Tilde{e} Q\chi\right)r^3 +  \left(a^2-\eta -\xi ^2+Q^2 (\tilde{e}^2-\chi ^2+1)\right) r^2\nonumber\\
   + &2  \left(a \tilde{e} Q (\xi -a \chi )+M \left((\xi -a \chi )^2+\eta \right)\right)r+Q^2 \left(2 a \xi  \chi-a^2 \chi ^2 -\eta -\xi ^2\right)-a^2 \eta.\label{KNR1}
\end{align}
The evaluation of $\int\frac{d\theta}{\sqrt{\Theta}}$ has been addressed in detail in the preceding section; here we focus exclusively on solving for $r$.

\subsection{null geodesic}
\begin{figure*}
\centering
  \includegraphics[width=.45\textwidth]{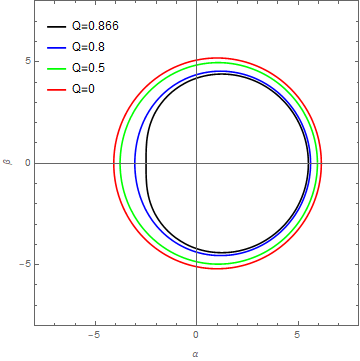}
\caption{The shadow of a Kerr-Newman black hole as seen by a distant observer in the equatorial plane. The mass of the black hole is set to $M=1$, and the spin parameter to $a=0.5$.}
\label{Fig:KerrNShad}
\end{figure*}

In the case of null geodesics, with $\tilde{e}=m=0$ and $\chi=1$, the function $R$ is further reduced to
\begin{equation}\label{RKN1}
    \frac{R}{E^2}=r^4+(a^2-\xi^2-\eta)r^2+2M[\eta+(\xi-a)^2]r-a^2\eta+Q^2 \left(2 a \xi-a^2 -\eta -\xi ^2\right).
\end{equation}
Solving $R=0$ and $R'=0$, we obtains
\begin{align}
\xi_c&= \frac{a^2 (M+r_c)+r_c \left(r_c (r_c-3 M)+2 Q^2\right)}{a (M-r_c)},\\
 \eta_c&=-\frac{r_c^2 \left(4 a^2 \left(Q^2-M r_c\right)+\left(r_c (r_c-3 M)+2 Q^2\right)^2\right)}{a^2 (M-r_c)^2}.  
\end{align}
As in the previous section, we compute the black hole shadow as seen by an observer in the equatorial plane. Figure \ref{Fig:KerrNShad} illustrates how the new parameter $Q$ affects the black hole shadow.

\begin{figure*}
\centering
\setlength{\tabcolsep}{20pt} 
  \renewcommand{\arraystretch}{1.5} 
  \begin{tabular}{cc} 
    \includegraphics[width=0.35\textwidth]{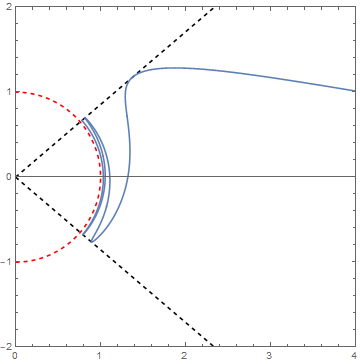} &
    \includegraphics[width=0.35\textwidth]{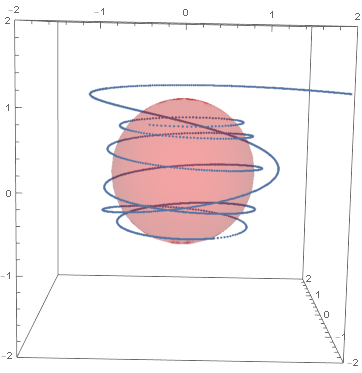} \\   
    \includegraphics[width=0.35\textwidth]{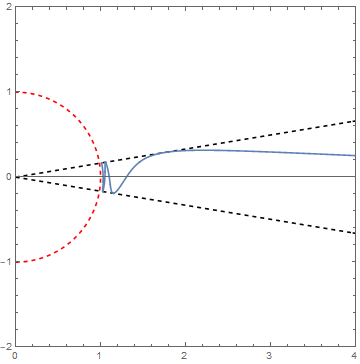} &
    \includegraphics[width=0.35\textwidth]{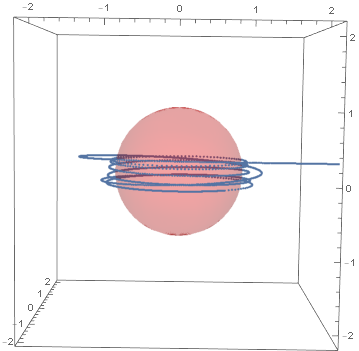} \\   
    \includegraphics[width=0.35\textwidth]{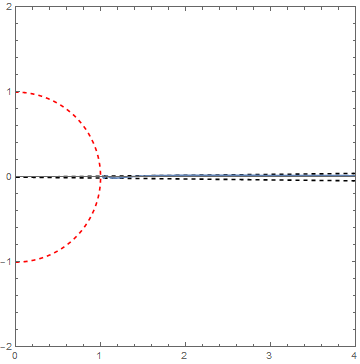} &
    \includegraphics[width=0.35\textwidth]{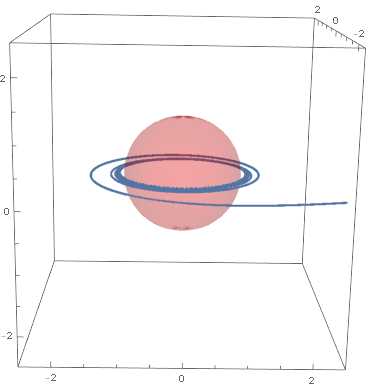} 
  \end{tabular}
\caption{Critical null geodesics with triply coincident roots in Kerr-Newman spacetime. The left diagram illustrates the trajectory of a null geodesic in the $(r,\theta)$ plane, while the right diagram shows the trajectory in three-dimensional space. The mass of the black hole is set to $M=1$, and the initial conditions are $r_{i}=10,\mu_{i}=0,\varphi_i=0$. The parameters for each row of images (from top to bottom) are $Q=0.5, Q=0.86,$ and $Q=0.866$. The red dashed line (ball) is located at $r=r_c$, and the black dashed lines are at $\mu=\pm \mu_{m}$.}
\label{Fig:KNNull1}
\end{figure*}

For critical orbits with triply coincident roots $R=0,R'=0,R''=0$, we obtain
\begin{align}
 \eta^{*}_c&=\frac{r_c^3 \left(3 M r_c-4 Q^2\right)}{3 M^2 r_c-M \left(Q^2+3 r_c^2\right)+r_c^3},\\
\xi^{*}_c&=\frac{3 M^2 r_c-M Q^2-r_c^3}{\sqrt{M} \sqrt{3 M^2 r_c-M \left(Q^2+3 r_c^2\right)+r_c^3}},\\
\frac{a^{*}_c}{M}&=\sqrt{\frac{r_c^3}{M^3}-3\frac{r_c^2}{M^2}+3\frac{r_c}{M}-\frac{Q^2}{M^2}}.
\end{align}
One can re-express $r_c=M-(M(M^2-a^2-Q^2 ) )^{1/3}$. Since the photon sphere lies outside the event horizon, with $r_+ \leq r_c$, we have $r_c = r_+ = M$. It follows that $\xi^{*}_c=\frac{M^2(3M^2-4Q^2)}{M^2-Q^2},\eta^{*}_c=\frac{2M^2-Q^2}{\sqrt{M^2-Q^2}},a^{*}_c=\sqrt{M^2-Q^2}$. Then, $\frac{R}{E^2}$ reduces to the expression (\ref{Rrd}), and
 \begin{align}
     \frac{\tilde{\Theta}}{E^2}=(M^2-Q^2)(\mu_1^2+\mu^2)(\mu_2^2-\mu^2),
 \end{align}
where $\mu_{1,2}^2=\frac{2M\sqrt{3M^2-Q^2}\pm 3M^2}{M^2-Q^2}$. Thus, the charge $Q$ affects the geodesic by altering $\theta_m$. The integral $\int \frac{dr}{\sqrt{R}}$ is given by Eq. (\ref{Rint1}), and $\int\frac{d\mu}{\sqrt{\tilde{\Theta}}}$ is given by Eq. (\ref{ThetaInt}) for the case of $m=0$. The numerical results for this case are shown in Fig. \ref{Fig:KNNull1}.

\begin{figure*}
\centering
\setlength{\tabcolsep}{20pt} 
  \renewcommand{\arraystretch}{1.5} 
  \begin{tabular}{cc} 
    \includegraphics[width=0.35\textwidth]{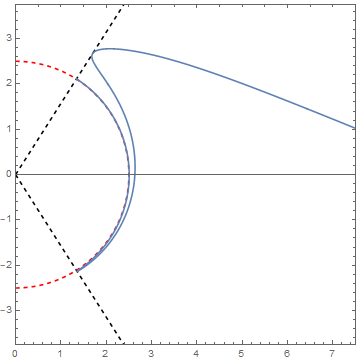} &
    \includegraphics[width=0.35\textwidth]{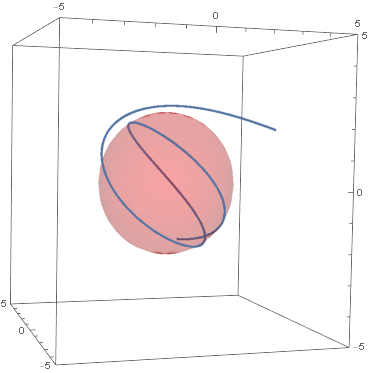} \\    
    \includegraphics[width=0.35\textwidth]{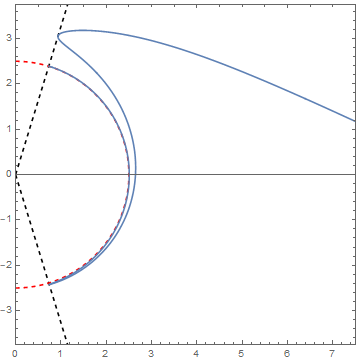} &
    \includegraphics[width=0.35\textwidth]{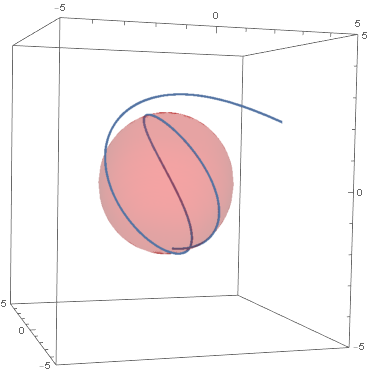} \\    
    \includegraphics[width=0.35\textwidth]{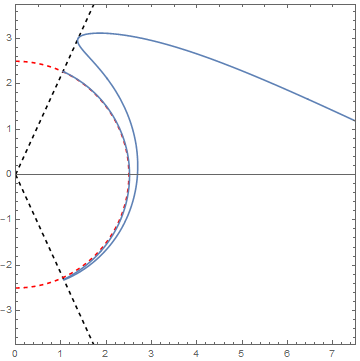} &
    \includegraphics[width=0.35\textwidth]{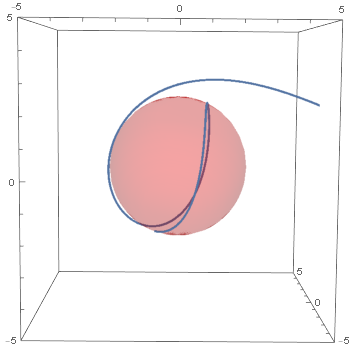} 
  \end{tabular}
\caption{Critical null orbits corresponding to double coincident roots in Kerr-Newman spacetime are shown. The left diagram illustrates the trajectory of a null geodesic in the $(r,\theta)$ plane, while the right diagram shows the trajectory of a null geodesic in three-dimensional space. The mass of the black hole is set to $M=1$, and the initial conditions are $r_{i}=10,\mu_{i}=0,\varphi_i=0$. The parameters for each row of images (from top to bottom) are $Q=0.3, Q=0.5,$ and $Q=0.866$ with $a=0.5,$ and $r_c=2.5$. The red dashed line (ball) marks $r=r_c$ and the black dashed line denotes $\mu=\pm \mu_{m}$.}
\label{Fig:KerrNull3}
\end{figure*}

In the case of coincident double roots, one substitutes $\xi=\xi_c,\eta=\eta_c$ into Eq. (\ref{RKN1}) to obtain
\begin{equation}
    \frac{R}{E^2}=(r-r_c)^2\left(r^2+2r_c r-\frac{a^2\eta_c}{r_c^2}-\frac{4Q^2\Delta(r_c)}{(r_c-M)^2}  \right).
\end{equation}
Defining $x=\frac{1}{r-r_c}$, one can also obtain the expression (\ref{Rint2}) with a different value of $c$, namely $c=3r_c^2-\frac{a^2\eta_c}{r_c^2}-\frac{4Q^2\Delta(r_c)}{(r_c-M)^2} $. Figure \ref{Fig:KerrNull3} illustrates the solutions for critical orbits with a double root for different values of $Q$ in the Kerr-Newman background.

\subsection{Timelike wordline}
For timelike critical orbits, analytical calculations become increasingly challenging. Owing to the additional parameters, the expressions involved in such computations are highly complex. We begin by examining the case of a triple root. Solving the system $R=0$, $R'=0$, and $R''=0$ directly is difficult, so we can only solve each equation individually. We can rewrite $R''=0$ as
\begin{equation}
   6 r^2- 6\left(\tilde{e} Q \chi -M \chi ^2+M\right) r+\left(a^2-\eta -\xi ^2+Q^2 \left(\tilde{e}^2-\chi ^2+1\right)\right)=0,
\end{equation}
and obtains
\begin{equation}
    \eta_c^{*}=6 r_c^2+\left(-6 \tilde{e} Q \chi +6 M \chi ^2-6 M\right)r_c+ \left(a^2+\tilde{e}^2 Q^2-\xi ^2-Q^2 \chi ^2+Q^2\right).
\end{equation}
Eliminating $\eta$ in $R'=0$, one obtains
\begin{align}
     &\left(-a^2 \tilde{e} Q \chi +a^2 M \chi ^2+a^2 M+\tilde{e}^2 M Q^2-M Q^2 \chi ^2+M Q^2\right)+r_c \left(-6 \tilde{e} M Q \chi +6 M^2 \chi ^2-6 M^2\right)\nonumber\\
     &+r_c^2 \left(3 \tilde{e} Q \chi -3 M \chi ^2+9 M\right)-4 r_c^3+ \xi  (a \tilde{e} Q-2 a M \chi )=0.   
\end{align}
Thus, we obtain
\begin{align}
  \xi_c^{*}&= \frac{1}{a (\tilde{e} Q-2 M \chi )}\bigg( \left(a^2 \tilde{e} Q \chi +a^2 (-M) \chi ^2-a^2 M-\tilde{e}^2 M Q^2+M Q^2 \chi ^2-M Q^2\right)\nonumber\\
  &+r_c \left(6 \tilde{e} M Q \chi -6 M^2 \chi ^2+6 M^2\right)+r_c^2 \left(-3 \tilde{e} Q \chi +3 M \chi ^2-9 M\right)+4 r_c^3   \bigg). 
\end{align}
Eliminating $\xi$ and $\eta$ from the equation $R=0$, yields a quartic algebraic equation in $\chi$
\begin{equation}\label{chiS}
a_0 + a_1 \chi + a_2 \chi^2 + a_3 \chi^3 + a_4 \chi^4 = 0,
\end{equation}
where the coefficients $a_0$ – $a_4$ are listed in Appendix B. In principle, by solving equation (\ref{chiS}), a complete set of parameter values $\eta=\eta_c^{*}$, $\xi=\xi_c^{*}$, and $\chi=\chi_{c}^{*}$ can be obtained so that the equation $R=0$ has a triple root. However, due to the complexity of these coefficients, it is practically impossible to derive an explicit expression for $\chi=\chi_c^{*}$. It is also difficult to determine which parameter values satisfy the physical constraints.
 
Fortunately, we can adopt an alternative approach to analyze the triple-root scenario. In fact, we know that the function $R$ can always be reduced to
\begin{equation}
  \frac{R}{E^2-m^2}=(r-r_c)^3(r-r_1).  
\end{equation}
By comparing with the expression (\ref{KNR1}), we find that $r_1=2 \tilde{e} Q \chi_c^{*} -2 M \chi_c^{*}{} ^2+2 M-3 r_c$, where $\chi=\chi_c^{*}$ satisfies Eq. (\ref{chiS}). The integration yields
\begin{equation}
    \frac{1}{\sqrt{E^2-m^2}}\int \frac{dr}{\sqrt{R}}=-\frac{2}{r_c-r_1}\sqrt{\frac{r-r_1}{r-r_c}}+C.
\end{equation}
It is worth noting that, for this scenario to occur, the parameters must still satisfy reasonable conditions, including the following: (1) the existence of a black hole horizon; (2) 
$r_c$ lying outside the horizon; (3) $\chi>1$; (4) 
$r_1<r_c$; (5) $\eta\geq 0$; and so on. We have not found a fully consistent set of parameter values, although we cannot entirely rule out this possibility.

\begin{figure*}
\centering
\setlength{\tabcolsep}{20pt} 
  \renewcommand{\arraystretch}{1.5} 
  \begin{tabular}{cc} 
    \includegraphics[width=0.35\textwidth]{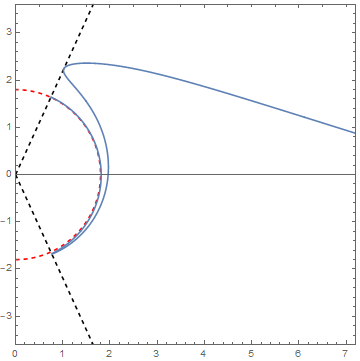} &
    \includegraphics[width=0.35\textwidth]{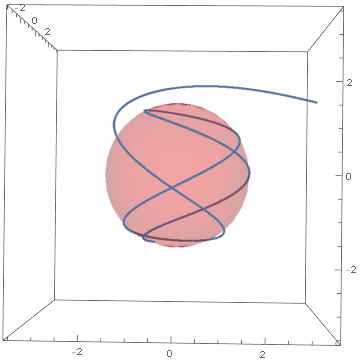} \\    
    \includegraphics[width=0.35\textwidth]{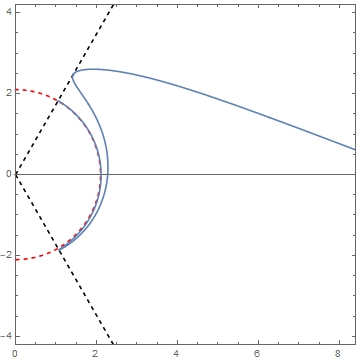} &
    \includegraphics[width=0.35\textwidth]{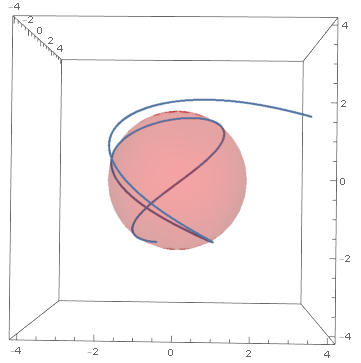} \\    
    \includegraphics[width=0.35\textwidth]{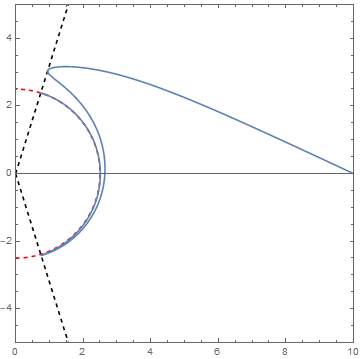} &
    \includegraphics[width=0.35\textwidth]{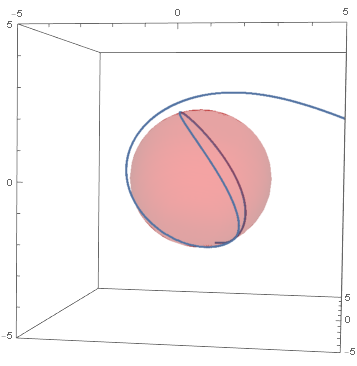} 
  \end{tabular}
\caption{Critical timelike orbits with coincident double roots in Kerr-Newman spacetime. The left diagram illustrates the trajectory of a timelike geodesic in the $(r,\theta)$ plane, while the right diagram shows the corresponding trajectory in three-dimensional space. The initial conditions are $r_{i}=10,\mu_{i}=0,\varphi_i=0$. The parameter sets for each row of images (from top to bottom) are $a=0.8, r_c=1.8, Q=0.5, \tilde{e}=1.1, \chi=1.1; a=0.8, r_c=2.1, Q=0.3, \tilde{e}=1.3, \chi=1.3;
a=0.4, r_c=2.5, Q=0.6, \tilde{e}=1.5, \chi=1.5$. }
\label{Fig:KNTimelike}
\end{figure*}

The case of double root is similar. The expressions for the solutions $\eta=\eta_c$, $\xi=\xi_c$ that satisfy the conditions $R=0$ and $R'=0$ are extremely complex, so we do not list them here. The function $R$ then degenerates into
\begin{equation}
    \frac{R}{E^2-m^2}=(r-r_c)^2\left(r^2+2(r_c+M(\chi^2-1)-\tilde{e}Q\chi)r+\frac{Q^2(2a\xi_c\chi-a^2\chi^2-\eta_c-\xi^2_c )-a^2\eta_c}{r_c^2} \right).
\end{equation}
The integral $\int\frac{dr}{\sqrt{E^2-m^2}\sqrt{R}}$ retains the form (\ref{RintKN1}), albeit with different values of the parameters $b$ and $c$:
\begin{align}
    b&=2r_c+M(\chi^2-1)-\tilde{e}Q\chi,\\
    c&=3r_c^2+r_c(M(\chi^2-1)-\tilde{e}Q\chi )+\frac{Q^2(2a\xi_c\chi-a^2\chi^2-\eta_c-\xi^2_c )-a^2\eta_c}{r_c^2} .
\end{align}
The numerical results are shown in Fig. \ref{Fig:KNTimelike}.

\section{Conclusions and Discussion}
When a test particle moves from infinity toward a black hole, orbits that neither fall into the black hole nor are scattered by it are referred to as critical orbits. This corresponds to the equation $R(r) = 0$ for the radial kinetic energy having either a double or a triple real root. This paper investigates all possible cases of critical orbits, including their motion in four different black hole backgrounds and for three types of particles (massless, massive uncharged, and massive charged). 

In spherically symmetric spacetimes, conservation of angular momentum confines orbits to a single plane. For critical orbits in the equatorial plane, the relation between the radius $r$ and the angle $\varphi$ can be expressed analytically. In axisymmetric spacetimes, however, the Carter constant supplements the conserved axial angular moentum, and generic orbits are three-dimensional. In the $\theta$-direction, the orbit oscillates about the equatorial plane. We provide analytical expressions for $(r, \theta, \varphi)$ and present the corresponding numerical results.

For null geodesics, the critical orbit lies at the boundary of  accreting light rays. Consequently, its shape determines the appearance of the black hole as seen from infinity. For a Schwarzschild black hole, this boundary is a circle of radius $3\sqrt{3}M$. If the black hole is charged, this radius decreases. Moreover, as the spin parameter $a$ increases, the shape becomes increasingly D-shaped.

In addition, we present a detailed discussion of the relationships between the physical parameters and the critical radius, which are crucial for modeling the accretion of collisionless Vlasov gas onto black holes. In future work, we will develop a more general theory of black hole accretion based on this study.

\appendix
\section*{Appendix A: the critical timelike geodesics with triple roots in Kerr background}
A triple root requires $R=0, R'=0$ and $R''=0$, which shows
\begin{align}
  r^4+2M(\chi^2-1)r^3+(a^2-\xi^2-\eta)r^2+2M[\eta+(\xi-a\chi)^2]r-a^2\eta&=0,\label{A1}\\
  2r^3+3M(\chi^2-1)r^2+(a^2-\xi^2-\eta)r+M[\eta+(\xi-a\chi)^2]&=0,\label{A2}\\
  6r^2+6M(\chi^2-1)r+(a^2-\xi^2-\eta)&=0.\label{A3}
\end{align}
Solving Eq. (\ref{A3}), one obtains
\begin{equation}\label{SolA1}
    \eta^{*}_c=a^2-\xi_c^2+6r_c^2+6M(\chi_c^2-1)r_c.
\end{equation}
Inserting the solution into Eq. (\ref{A2}), one obtains
\begin{equation}
    a^2 M \chi_c ^2+a^2 M+6 M^2 r_c \chi_c ^2-6 M^2 r_c-3 M r_c^2 \chi_c ^2+9 M r_c^2-4 r_c^3- 2\xi_c  a M \chi_c =0.
\end{equation}
Then, the solution of $\xi_c$ is given by
\begin{equation}\label{SolA2}
   \xi^{*}_c= \frac{a^2 M \chi_c ^2+a^2 M+6 M^2 r_c \chi_c ^2-6 M^2 r_c-3 M r_c^2 \chi_c ^2+9 M r_c^2-4 r_c^3}{2 a M \chi_c }.
\end{equation}
Inserting the expressions (\ref{SolA1}) and (\ref{SolA2}) into $R$ and defining: 
\begin{align}
   A&=  (a^2M-r_c(6M^2-9Mr_c+4r_c^2 ))^2,\\
  B&=a^4 M+2 a^2 r_c \left(-6 M^2+3 M r_c+2 r_c^2\right)+r_c^2 \left(36 M^3-68 M^2 r_c+45 M r_c^2-12 r_c^3\right),\\
  C&=a^4-6 a^2 r_c (2 M+r_c)+r_c^2 \left(36 M^2-28 M r_c+9 r_c^2\right),
\end{align}
we can reexpress $R=0$ as
\begin{align}
    \frac{1}{4 M^2 \chi_c ^2}(M^2C\chi_c^4 -2MB\chi_c^2+A)=0.   
\end{align}
Thus, we have
 \begin{equation}
     \chi_c^2=\frac{B\pm\sqrt{B^2-AC}}{MC},
 \end{equation}
where $B^2-AC=16 M r_c^3 \Delta^3(r_c)$. Over the interval $r_c\in[r_+,\infty)$, the function $f(r_c)=B+\sqrt{B^2-AC}$ is monotonically decreasing, whereas $C(r_c)$ is monotonically increasing. For $a\in[0,M]$, one can show that $f(r_+)\leq0$ and $C(r_+)\geq 0$. Thus, for $r_c> r_+$, we have $\chi_c^2<0$, and no physically admissible parameter range exists.
\section*{Appendix B: the coefficients $a_0$–$a_4$}
\begin{align}
    a_0&=a^2\bigg( \left(a^2+\left(\tilde{e}^2+1\right) Q^2\right) \left(a^2 (\tilde{e} Q+M) (M-\tilde{e} Q)+\left(\tilde{e}^2+1\right) M^2 Q^2-\tilde{e}^2 Q^4\right) \nonumber\\
    &+6r_c\big(a^2 \left(\tilde{e}^2 M Q^2-2 M^3\right)-2 \left(\tilde{e}^2+1\right) M^3 Q^2+\tilde{e}^2 M Q^4 \big) -6 r_c^2\big(a^2 \left(\tilde{e}^2 Q^2-3 M^2\right)\nonumber\\
    &-3 \left(\tilde{e}^2+1\right) M^2 Q^2+\tilde{e}^2 Q^4-6 M^4 \big)-2 r_c^3 \left(M \left(4 a^2+\left(5 \tilde{e}^2+4\right) Q^2+54 M^2\right)\right)\nonumber\\
    &+3 r_c^4 \left(\tilde{e}^2 Q^2+43 M^2\right)-72 M r_c^5+16 r_c^6\bigg),\\
    a_1&=a^2 \tilde{e} Q\bigg(2 M \left(a^2+Q^2\right) \left(a^2+\left(\tilde{e}^2+1\right) Q^2\right) +6r_c\big(-2 (2 + en^2) M^2 Q^2 + en^2 Q^4\nonumber\\
    & + a^2 (-4 M^2 + en^2 Q^2)  \big)+6M r_c^2\big(2 a^2+\left(\tilde{e}^2+2\right) Q^2+12 M^2 \big)   +   r_c^3 \big(8 a^2-2 \left(\tilde{e}^2-4\right) Q^2\nonumber\\
    & -136 M^2\big)+90 M r_c^4-24 r_c^5\bigg),
\end{align}
\begin{align}
    a_2&=a^2\bigg( \left(a^2+Q^2\right) \left(a^2 \left(\tilde{e}^2 Q^2-2 M^2\right)-2 \left(\tilde{e}^2+1\right) M^2 Q^2+\tilde{e}^2 Q^4\right)
 \nonumber\\
 &+6r_c \big( a^2 \left(4 M^3-3 \tilde{e}^2 M Q^2\right)+2 \left(\tilde{e}^2+2\right) M^3 Q^2-3 \tilde{e}^2 M Q^4\big) \nonumber\\
 &+r_c^2\big(-6 a^2 \left(\tilde{e}^2 Q^2+2 M^2\right)+6 \left(5 \tilde{e}^2-2\right) M^2 Q^2-6 \tilde{e}^2 Q^4-72 M^4 \big)\nonumber\\
 &+r_c^3 \left(-8 a^2 M-2 \left(13 \tilde{e}^2+4\right) M Q^2+136 M^3\right)+r_c^4 \left(9 \tilde{e}^2 Q^2-90 M^2\right)+24 M r_c^5
 \bigg),\\
    a_3&=2M a^2 \tilde{e} Q\bigg( -\left(a^2+Q^2\right)^2 +12 M (a^2 + Q^2) r_c+6 r_c^2 \left(a^2-6 M^2+Q^2\right) +28 M r_c^3-9 r_c^4 \bigg),\\
    a_4&=M^2 a^2\bigg(\left(a^2+Q^2\right)^2-12 r_c \left(M \left(a^2+Q^2\right)\right) -6 r_c^2 \left(a^2-6 M^2+Q^2\right) -28 M r_c^3+ 9 r_c^4\bigg).
\end{align}

\textbf{Acknowledgement}
This work was supported in part by the National Natural Science Foundation of China (Grant No. 12505071) and by the Research Foundation of the Education Bureau of Hunan Province, China (Grant No. 25B0635). It was also supported in part by the Key Laboratory of Information Detection and Intelligent Processing Technology of the Hunan Provincial Department of Education, and by the Applied Characteristic Subject of Hunan Province, "Electronic Science and Technology".


\end{document}